\documentstyle[mncite,psfig]{mn}

\def\gte{\,\lower.6ex\hbox{$\buildrel >\over \sim$} \, }
\def\e{\,\lower.6ex\hbox{$\buildrel <\over \sim$} \, }

\title{The Statistical Analysis of Star Clusters}

\author[A. Cartwright]{Annabel Cartwright, Anthony P. Whitworth\\
            Department of Physics and Astronomy, Cardiff University\\
          }

\date{}

\begin{document}

\maketitle

\begin{abstract}

We review a range of stastistical methods for analyzing the structures of 
star clusters, and derive a new measure ${\cal Q}$ which both quantifies, 
and distinguishes between, a (relatively smooth) large-scale radial density 
gradient and multi-scale (fractal) sub-clustering.

The distribution of separations $p(s)$ is considered, and the Normalised 
Correlation Length $\bar{s}$ (i.e. the mean separation between stars, divided 
by the overall radius of the cluster), is shown to be a robust indicator of 
the extent to which a smooth cluster is centrally concentrated. For spherical 
clusters having volume-density $n \propto r^{-\alpha}$ (with $\alpha$ between 
0 and 2) $\bar{s}$ decreases monotonically with $\alpha$, from $\sim 0.8$ to 
$\sim 0.6$. Since $\bar{s}$ reflects all star positions, it implicitly 
incorporates edge effects. However, for fractal star clusters (with fractal 
dimension $D$ between 1.5 and 3) $\bar{s}$ decreases monotonically with 
$D$ (from $\sim 0.8$ to $\sim 0.6$). Hence $\bar{s}$, on its own, can 
quantify, but cannot distinguish between, a smooth large-scale radial 
density gradient and multi-scale (fractal) sub-clustering.

The Minimal Spanning Tree (MST) is then considered, and it is shown that the 
Normalised Mean Edge Length $\bar{m}$ (i.e. the mean length of the branches 
of the tree, divided by $({\cal N}_{\rm total}A)^{1/2}/({\cal N}_{\rm total}-1)$, 
where $A$ is the area of the cluster and ${\cal N}_{\rm total}$ is the number
of stars), can also quantify, but again cannot on its own distinguish between, 
a smooth large-scale radial density gradient and multi-scale (fractal) sub-clustering.

However, the combination ${\cal Q} ={ \bar{m}}/{ \bar{s}} $ does both 
quantify, and distinguish between, a smooth large-scale radial density 
gradient and multi-scale (fractal) sub-clustering. IC348 has 
${\cal Q} = 0.98$ and $\rho$ Ophiuchus has ${\cal Q} = 0.85$, implying 
that both are centrally concentrated clusters with, respectively, 
$\alpha \simeq 2.2 \pm 0.2$ and $\alpha \simeq 1.2 \pm 0.3\,$. 
Chamaeleon and IC2391 have ${\cal Q} =0.67$ and ${\cal Q} =0.66$ 
respectively, implying mild substructure with a notional fractal 
dimension $D \simeq 2.25 \pm 0.25$. Taurus has 
even more sub-structure, with ${\cal Q} =0.45$ implying $D' \simeq 1.55 
\pm 0.25$. If the binaries in Taurus are treated as single systems, ${\cal Q} $
increases to $0.58$ and $D'$ increases to $1.9 \pm 0.2$.
\end{abstract}


\section{Introduction}

Since most stars are formed in clusters, it would be useful to have 
quantitative and objective statistical measures of their structure, 
with a view to comparing clusters formed in different environments, 
and tracking changes in structure as clusters evolve. This is 
particularly important for young, embedded clusters, where the structure 
may yield important clues to the formation process but is changing 
rapidly. It is also important for comparing observed clusters with 
numerical simulations.

At present, we do not have sufficiently robust statistical measures 
for this purpose. Features which are easily identified by the human 
eye, such as sub-clusters, or linear features, can be strangely elusive 
to objective statistical analysis. For example, it is difficult to distinguish,
statistically, between a degree of fractal or random sub-clustering, and 
the existence of a density gradient (Bate, Clarke \& McCaughrean 
1997). This paper explores some possible measures, and evaluates their 
usefulness. In particular, we find a robust objective measure which 
both quantifies, and distinguishes between, a smooth large-scale radial 
density gradient and multi-scale (fractal) sub-clustering.

In Section 2 we describe our methodology. In Section 3 we look again at 
the Mean Surface Density of Companions (MSDC), a tool pioneered by Larson 
(1995) and subsequently used by several others (e.g. Simon 1995; Bate, 
Clarke and McCaughrean 1997; Nakajima et al. 1998; Brandner \& K\"ohler 
1998; Gladwin et al. 1999; Klessen \& Kroupa 2001). We focus on measures 
which reflect the clustering regime (wide separations) rather than the 
binary regime (close separations). In Section 4 we explore the use of 
the Minimal Spanning Tree (Barrow, Bhavasar \& Sonoda 1985) and its 
derivatives. In Section 5 we combine the MSDC and the MST to derive a 
single measure ${\cal Q}$ which is able both to quantify, and to 
distinguish between, a smooth radial density gradient and multi-scale 
(fractal)  sub-clustering.

All the measures are tested and calibrated on multiple realizations 
of artificial star clusters, and applied to $\rho$ Ophiuchus, Chamaeleon, 
Taurus, IC348 and IC2391. Our results are discussed in Section 6, and 
the main conclusions are summarized in Section 7.

\section {Methodology}

Three different types of artificial star cluster have been created, using 
random numbers ${\cal R}$ to generate the individual star positions. The 
first type (2D$\alpha$) are circular clusters (i.e. two-dimensional disc-like 
clusters) with surface density $N \propto r^{-\alpha}$ and 
$\alpha = 0 \;{\rm or}\; 1$. The second type (3D$\alpha$) are spherical 
clusters (i.e. three-dimensional clusters) having volume density 
$n \propto r^{-\alpha}$ with $\alpha = 0,\;1,\;2,\;{\rm and}\;2.9$. The 
third type (F$D$) are fractal star clusters (again three dimensional) 
with fractal dimension $D = 3.0,\;2.5,\;2.0,\;{\rm or}\;1.5$. 

The different types are listed in Column 1 of Table~\ref{Measures}. All of 
the artificial clusters are created with 100 to 300 stars, as the numbers 
of stars within the  five real clusters lie within that range. The data 
for the five real clusters used are illustrated in  Appendix~\ref{RawData}, 
and the sources listed in Table 2.


\begin {table}

\caption{Clustering measures obtained for artificial and real star 
clusters. Column 1 lists the cluster type (for artificial clusters) 
or name (for real clusters). Column 2 gives the Normalized Correlation 
Length $\bar{s}$ (i.e. the ratio of the mean separation to the cluster 
radius, see Section 3). Column 3 gives the Normalised Mean Edge Length 
$\bar{m}$ (see Section 4). Column 4 gives the mean value of the standard 
deviation of the edge length, $\bar{\sigma}_m$. Column 5 gives 
${\cal Q} = \bar{m} / \bar{s}$. For the artificial star clusters, 
means and standard deviations are computed from 100 realisations of 
each type, with $100 \leq {\cal N}_{total} \leq 300$.}
\begin {tabular}  [tbp]{lcccc} 
Cluster type  & & &\\
or name&  $\bar{s}$ & $\bar{m}$ & $\bar{\sigma}_m$ & ${\cal Q}$\\ \hline
 & &  & & \\
2D0($N\!\propto\!r^0$)	 & $.88\pm.03$ &$.65 \pm0.02$ &$.31\pm.02$&$.74\pm.02$\\
2D1($N\!\propto\!r^{-1}$) & $.70\pm.03$ &$.61 \pm.02 $ &$.38\pm.02$&$.85\pm.03$\\

 & &  & & \\
3D2.9($n\!\propto\!r^{-2.9}$) &$.16\pm.02$&$.24\pm.05$& $.59\pm.07$&$1.50\pm.13$\\
3D2($n\!\propto\!r^{-2}$) & $.60\pm.03$ &$.55 \pm.02 $& $.41\pm.03$&$.93\pm.03$\\
3D1($n\!\propto\!r^{-1}$) & $.73\pm.03$ &$.61 \pm.02 $& $.33\pm.03$&$.84\pm.02$\\
3D0($n\!\propto\!r^0$) 	 & $.80\pm.02$ &$.63\pm.02  $& $.31\pm.02$&$.79\pm.02$\\
 & & & & \\
F3.0($D\!=\!3.0$)& $.81\pm.03$ 	& $.64 \pm.02$ 	&$ .30\pm.02$	&$.80\pm.02$\\
F2.5($D\!=\!2.5$)& $.74\pm.09$ 	& $.54 \pm.05$ 	&$ .28\pm.03$	&$.73\pm.06$\\
F2.0($D\!=\!2.0$)& $.67\pm.13$ 	& $.41 \pm.04$ 	&$ .28\pm.02$	&$.61\pm.08$\\
F1.5($D\!=\!1.5$)& $.62\pm.18$ 	& $.27 \pm.07$ 	&$ .35\pm.07$	&$.45\pm.09$\\
 & & & &\\
IC2391		& 0.74 		& .49 		& .30 		& .66\\
Chamaeleon	& 0.63 		& .42 		& .45 		& .67\\
Taurus		& 0.55   	& .26 		& .56 		& .47\\
$\rho$ Ophiuchus  & 0.53 	& .45 		& .39 		& .85\\
IC348		& 0.49 		& .48 		& .46 		& .98\\  \hline
\end {tabular}
\label{Measures}

\end{table}


A cluster of type 2D$\alpha$ is created by positioning the stars according 
to
\begin{equation} \left. \begin{array}{lll}
r & = & \left\{(2-\alpha){\cal R}_r/2 \right\}^{1/(2-\alpha)} \,, \\
\phi & = & 2 \pi {\cal R}_\phi \,, \\
x & = & r \, {\rm cos}(\phi) \,, \\
y & = & r \, {\rm sin}(\phi) \,. \\
\end{array} \right\}
\end{equation}
where ${\cal R}_r$ and ${\cal R}_\phi$ are random numbers in the range 0-1.

A cluster of type 3D$\alpha$ is created by positioning the stars according 
to
\begin{equation} \left. \begin{array}{lll}
r & = & \left\{(3-\alpha){\cal R}_r/3 \right\}^{1/(3-\alpha)} \,, \\
\theta & = & {\rm cos}^{-1} \left( 2 {\cal R}_\theta - 1 \right) \,, \\
\phi & = & 2 \pi {\cal R}_\phi \,, \\
x & = & r \, {\rm sin}(\theta) \, {\rm cos}(\phi) \,, \\
y & = & r \, {\rm sin}(\theta) \, {\rm sin}(\phi) \,, \\
z & = & r \, {\rm cos}(\theta) \,. \\
\end{array} \right\}
\end{equation}
where ${\cal R}_r$, ${\cal R}_\theta$ and ${\cal R}_\phi$ are random numbers
in the range 0-1.
Clearly this method cannot be used for $\alpha$ = 3, so to have a cluster 
type approximating to $\alpha$ = 3 we use $\alpha$ = 2.9.

A cluster of type F$D$ is created by defining an ur-cube with side 2, and 
placing an ur-parent at the centre of the ur-cube. Next, the ur-cube is 
divided into ${\cal N}_{\rm div}^3$ equal sub-cubes, and a child is placed 
at the centre of each sub-cube (the first generation). Normally we use 
${\cal N}_{\rm div} = 2$, in which case there are 8 sub-cubes and 8 
first-generation children. The probability that a child matures to become 
a parent in its own right is ${\cal N}_{\rm div}^{(D-3)}$, where $D$ is 
the fractal dimension. For lower $D$, the probability that a child matures 
to become a parent is lower, and the cluster is more `porous'. Children 
who do not mature are deleted, along with the ur-parent. A little noise 
is then added to the positions of the remaining children, to avoid an 
obviously regular structure, and they then become the parents of the next 
generation, each one spawning ${\cal N}_{\rm div}^3$ children (the second 
generation) at the centres of ${\cal N}_{\rm div}^3$ equal-volume 
sub-sub-cubes, and with each second-generation child having a probability 
${\cal N}_{\rm div}^{(D-3)}$ of maturing to become a parent. This process 
is repeated recursively until there is a sufficiently large generation 
that, even after pruning to impose a spherically symmetric envelope of 
radius 1 within the ur-cube, there are still more children than the 
required number of stars. Children are then culled randomly until the 
required number is left, and the surviving children are identified with 
the stars of the cluster. At each generation, the survival of a child is 
determined by generating a random number ${\cal R}$ in $(0,1)$; 
survival then requires that ${\cal R} < {\cal N}_{\rm div}^{(D-3)}$.

Clusters of type 2D$\alpha$ are investigated for two purposes. First, 
we wish to clarify the effect of a sharply defined circular edge on 
an otherwise statistically uniform, two-dimensional distribution of 
stars. Clusters of type 2D0 enable us to isolate this effect. Second, 
we wish to explore how readily two-dimensional and three-dimensional 
distributions can be distinguished. This could be important if stars 
are being formed in layers, for example at a shock front.

For each type of artificial cluster, 100 realisations are analysed, 
so that means and standard deviations can be obtained for the 
parameters extracted. Three-dimensional clusters (types 3D$\alpha$ 
and F$D$) are projected onto an arbitrary plane prior to analysis. 
Two-dimensional clusters are viewed face-on.


\begin{figure*}
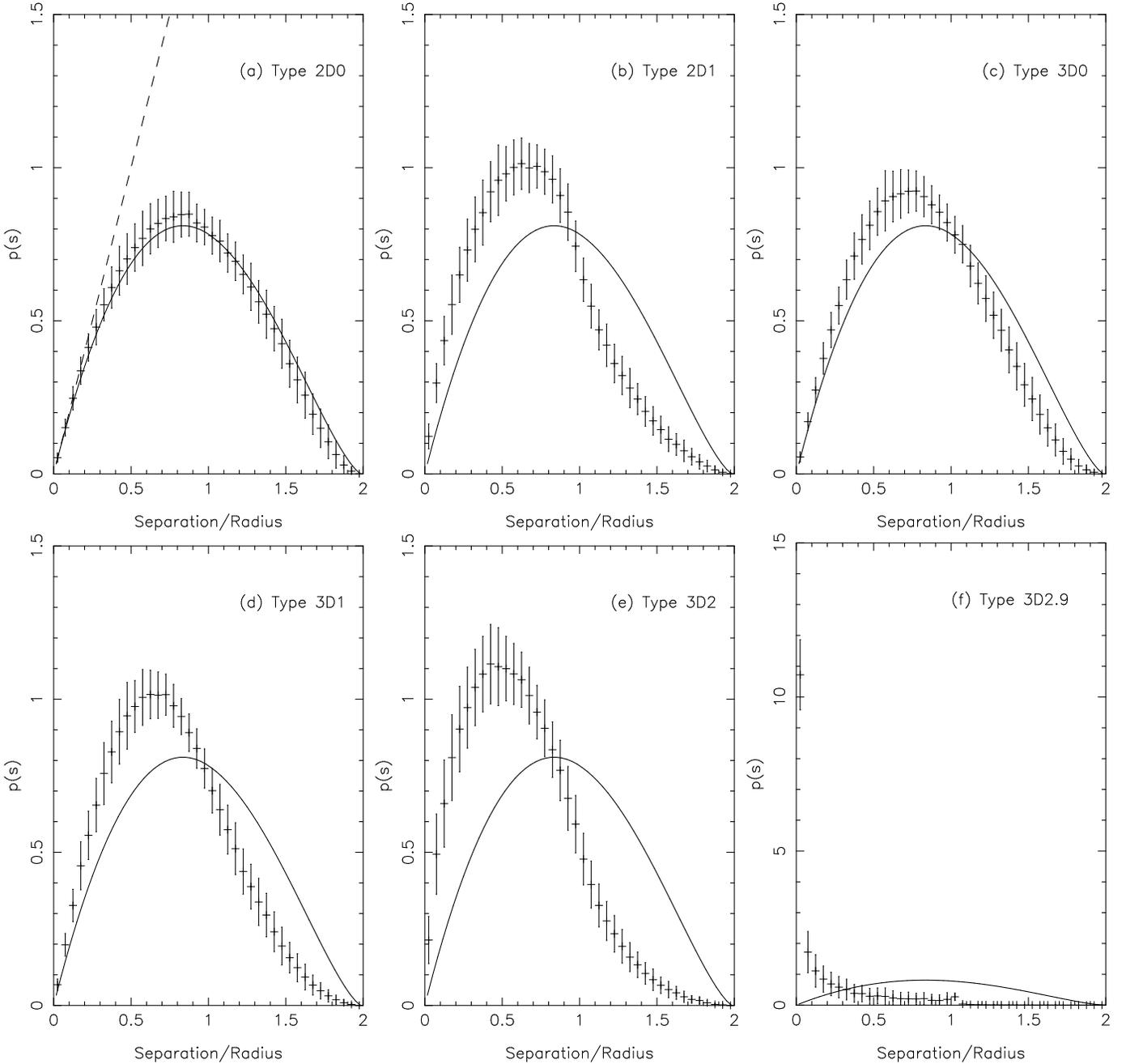

\centerline{\psfig{figure=fig2a.ps,width=6cm}
            \psfig{figure=fig2b.ps,width=6cm}
            \psfig{figure=fig2c.ps,width=6cm}}

\centerline{\psfig{figure=fig2d.ps,width=6cm}
            \psfig{figure=fig2e.ps,width=6cm}
            \psfig{figure=fig2f.ps,width=6cm}}
\caption{Distribution function $p(s)$ for separations between randomly 
chosen stars in artificial (non-fractal) cluster of type 
(a) 2D0, $N \propto r^0$;
(b) 2D1, $N \propto r^{-1}$; 
(c) 3D0, $n \propto r^0$; 
(d) 3D1, $n \propto r^{-1}$; 
(e) 3D2, $n \propto r^{-2}$; and 
(f) 3D2.9, $n \propto r^{-2.9}$. The solid line is the value of $p(s)$ 
for a star cluster of type 2D0 having an infinite number of stars (Eqn. 
\ref{edge-effects}), and is included for reference. The dashed line is 
$p(s) = 2s$ (see text). $s$ is normalized to the overall radius of the 
cluster, as described in the text.}
\label{Arti/p(s)}
\end{figure*}


 
\begin{figure*}
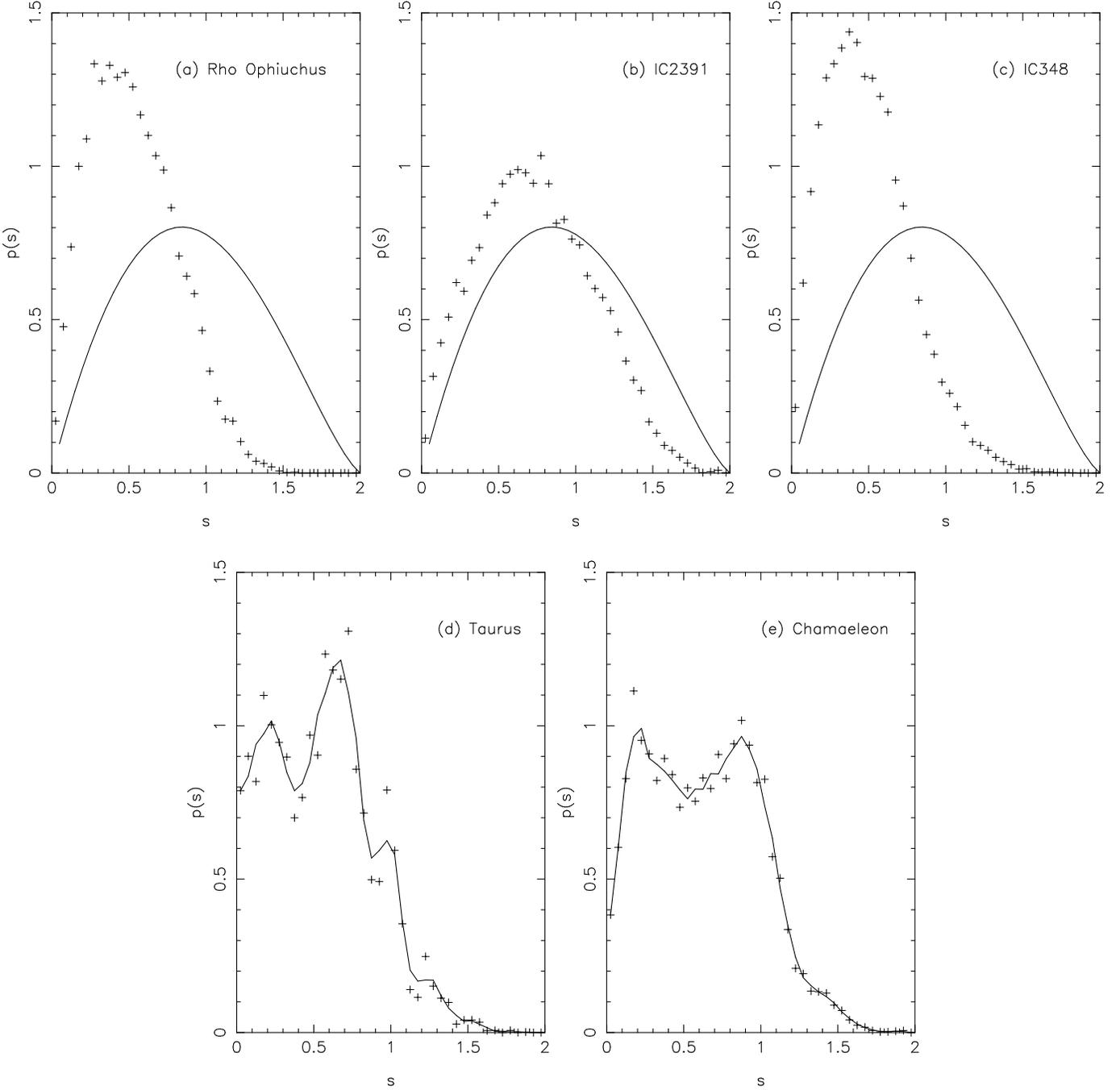

\centerline{\psfig{figure=fig3a.ps,width=6cm}
            \psfig{figure=fig3b.ps,width=6cm}
            \psfig{figure=fig3c.ps,width=6cm}}
\vspace{0.5cm}
\centerline{\psfig{figure=fig4a.ps,width=6cm}
            \psfig{figure=fig4b.ps,width=6cm}}
\caption{Distribution functions for separations between randomly chosen 
stars in five real star clusters. (a) $\rho$ Ophiuchus, (b) IC2391, (c) 
IC348, (d) Taurus, and (e) Chamaeleon. The solid line is the value of 
$p(s)$ for a star cluster of type 2D0 having an infinite number of stars 
(Eqn. \ref{edge-effects}), and is included for reference. In (d) and (e), 
the solid line represents a smoothed version of the raw data, to show the 
existence of multiple maxima.}
\label{Real/p(s)}
\end{figure*}


\section{The Mean Surface Density of Companions}

\subsection{Log-log plots and edge effects}

A widely used tool for analysing the structure of star clusters 
is the log/log plot of the mean surface-density of companions, $\bar{N}$ 
against separation, $s$. This tool has been pioneered by Larson (1995), 
building on earlier work by Gomez et al. (1993), who  used the two 
point correlation function. Several papers have confirmed Larson's 
finding that a plot of $\ell og[\bar{N}]$ against $\ell og[s]$ --- 
hereafter a Larson Plot --- can be fitted with two power law sections, 
corresponding to two distinct regimes. At the smaller separations, 
$s < s_{\rm break}$, a star's companions are mainly in binary and 
higher multiple systems, and the slope of the Larson Plot is 
$\eta_{\rm binary} \equiv d\ell og[\bar{N}]/d\ell og[s] \simeq -2$. At 
larger separations, $s > s_{\rm break}$, companions are simply other 
members of the overall cluster, and may only be close due to projection. 
The slope here is generally larger (i.e. still negative but smaller in 
magnitude), $\eta_{\rm cluster} \equiv d\ell og[\bar{N}]/d\ell og[s] \ga -1$. 
Larson has suggested that $\eta_{\rm cluster}$ might be related to the fractal 
dimension of the sub-clustering, $D = \eta_{\rm cluster} + 2$. In addition, 
he has proposed that the break point between the two straight sections, 
at $s_{\rm break}$, might correspond to the Jeans length. However, recent 
analysis has cast some doubt on these interpretations. First, the break 
point is strongly influenced by the overall surface-density of stars (and 
hence by the depth of the cluster along the line of sight), as pointed 
out by Simon (1997) and Bate et al (1997). Second, fitting $\eta_{\rm 
cluster}$ objectively is difficult, because at the low-$s$ end it is 
distorted by the switch to the binary regime, and --- more importantly 
--- at the high-$s$ end it is distorted by edge effects. Consequently, 
one is left with at best a range of order $2 s_{\rm break}$ to 
$0.1 R_{\rm cluster}$  and the result is sensitive to how the range is 
actually chosen; if the range is shortened or extended arbitrarily, the 
slope of the fitted line may change dramatically. Third, 
$\eta_{\rm cluster}$ is not necessarily related to the fractal dimension 
of the clustering. As shown by Bate et al. (1997) and Klessen \& Kroupa 
(2001), it may simply reflect a large-scale density gradient in the cluster.

\subsection {Linear plots and edge effects}

An alternative way of evaluating the data from which Larson plots are 
derived is to calculate the distribution function $p(s)$, where $p(s)ds$ 
gives the probability that the projected separation between two cluster 
stars chosen at random is in the interval $(s,s+ds)$. To do this 
empirically, we define $i_{\rm max}$ equal $s$-bins in the range $\,0 < s < 
2 R_{\rm cluster}\,$, so that all the bins have width 
$\Delta s = 2 R_{\rm cluster}/ i_{\rm max}$, and the $i$th bin corresponds 
to the interval $\,(i-1) \Delta s < s < i \Delta s\,$, with mean value 
$s_i = (i - 1/2) \Delta s$. $R_{\rm cluster}$ is the overall radius of the 
cluster, and is defined by finding the mean position of all the stars 
in the cluster and then setting $R_{\rm cluster}$ equal to the distance to the 
furthest star. Then we count the number of separations ${\cal N}_i$ falling 
in each bin, and put
\begin{eqnarray}
p(s_i) & = & \frac{2\,{\cal N}_i}{{\cal N}_{\rm total}\,
({\cal N}_{\rm total}-1)\,\Delta s} \,,
\end{eqnarray}
where ${\cal N}_{\rm total}$ is the total number of stars in the cluster, 
and hence ${\cal N}_{\rm total}\,({\cal N}_{\rm total}-1)/2$ is the total 
number of separations. 

Figure~\ref{Arti/p(s)}(a) presents the results obtained from 100 clusters 
of type 2D0, i.e. a disc having statistically uniform surface-density. The 
plotted points give the mean $\bar{p}(s_i)$ from the 100 realizations, 
and the error bars give the width of the bin and the $1 \sigma$ 
standard deviation. If there were no edge effects (i.e. if the uniform 
surface-density extended to infinity in two dimensions), we would 
have $\bar{p}(s) = 2 s$, and this is indeed a good fit to $\bar{p}(s_i)$ at 
small $s_i$ values, as indicated by the dashed line on Fig.~\ref{Arti/p(s)}(a). 
{\em Departures from this straight line are entirely due to edge
 effects.}

In fact, $\bar{p}(s)$ can be calculated semi-analytically for a disc having 
uniform surface-density:
\begin{eqnarray}
\label{edge-effects}
\bar{p}(s) & = & \left\{ \begin{array} {ll}
2 s (1-s)^2 + \frac{4s}{\pi} \int_{1-s}^{1} \theta r dr \,, & 0 \leq s < 1 \,; \\
 & \\
\frac{4s}{\pi} \int_{s-1}^{1} \theta r dr \,, & 1 \leq s < 2 \,; \\
 & \\
0 \,, & s \geq 2 \,;
\end{array} \right.
\end{eqnarray}
where
\begin{eqnarray}
\theta & = & {\rm cos}^{-1}\left[ \frac{r^2 + s^2 - 1}{2 r s} \right]
\end{eqnarray}
The solid line on Fig.~\ref{Arti/p(s)}(a) shows that this function fits 
the plotted points well, and it is included on all the other plots for 
reference, i.e. to emphasize the features which are not due to edge effects.

When derived in this way, the $\bar{p}(s)$ plot contains little information 
about the distribution of binary separations, since they are all in the 
first bin. However, it seems to be well established that the 
distribution of binary separations is approximately scale free 
over a wide range of separations ($\eta_{\rm binary} \simeq -2$). The more 
critical issue --- the one with which we 
are concerned here --- is the distribution of separations in the 
clustering regime and what it tells us about the overall structure 
of the cluster. This information is well represented by $\bar{p}(s)$, as 
can be seen from Figs.~\ref{Arti/p(s)}(b) through ~\ref{Arti/p(s)}(f), 
which show the results obtained for the other five types of non-fractal 
artificial star cluster. Figure~\ref{Arti/p(s)}(b) shows how $\bar{p}(s)$ 
is slewed towards smaller $s$ values for a disc with a centrally 
concentrated surface-density, $N \propto r^{-1}$. Figures~\ref{Arti/p(s)}(c) 
to ~\ref{Arti/p(s)}(f) show spherical clusters having volume-density 
gradients $n \propto r^{-\alpha}$ with $\alpha = 0,\, 1,\, 2,\;{\rm and}\;2.9$. 
Again the distribution slews to smaller $s$ values as the sphere becomes 
more centrally concentrated (i.e. with increasing $\alpha$).

\subsection {The Normalised Correlation Length}
One feature which distinguishes the plots is the location of the maximum, 
i.e. the separation $s_{\rm max}$ at which $\bar{p}(s)$ is largest. As a 
cluster becomes more centrally condensed, $s_{\rm max}$ moves to 
smaller values, and the amplitude of the maximum increases. However, 
for an individual cluster $s_{\rm max}$ will not be well defined, and 
so it is not a robust measure.

A better measure of this trend is the Normalized Correlation Length 
for each cluster. The Correlation Length is the mean separation $\bar{s}$ 
between stars in the cluster, and it is normalized by dividing by 
$R_{\rm cluster}$. The second column of Table~\ref{Measures} gives mean 
values of $\bar{s}$ and their standard deviations, for the various 
artificial cluster types. The $\bar{s}$ values for the five real 
star clusters are also given.

The shapes of the $p(s)$ plots, and hence also the  $\bar{s}$ values, are
independent of the number of stars in the cluster. In trials 
with cluster sizes of 100 to 1000 stars, $\bar{s}$ stays within one 
standard deviation of the mean value for 200 stars. This is at first 
sight surprising. A 1000-star cluster is so much more dense than a 
100-star cluster, that one might expect the mean separation of stars to
 be smaller. However, although each star has more close neighbours, it
also has more distant neighbours, and the value of $\bar{s}$ remains
constant. This is an attractive feature of the Normalised Correlation 
Length as a statistical descriptor for clusters. From Table 1, we see 
that  $\bar{s}$ decreases monotonically with increasing $\alpha$, and 
can therefore be used to estimate $\alpha$ for star clusters which are 
presumed {\it a priori} to have radial density gradients.

Importantly, cluster types 2D1 and 3D2 are easily distinguished by 
their $\bar{s}$ values and their $\bar{p}(s)$ plots, despite the 
widespread but fallacious assumption that a three 
dimensional cluster with volume-density $n \propto r^{-2}$ is, when 
projected on the sky, similar to a two dimensional cluster with 
surface-density $N \propto r^{-1}$. In fact it is clusters of types 
2D1 and 3D1 (i.e. with the same exponent, $d \ell n[N]/d \ell n[r] 
\sim -1$, and $d \ell n[n]/d \ell n[r] \sim -1$) which are hard to 
distinguish.

$p(s)$ plots for the real clusters are shown on Fig.~\ref{Real/p(s)}. 
IC348 and $\rho$ Ophiuchus resemble clusters of type 3D2, both on the 
basis of their $\bar{s}$ values (Table 1), and the shapes of their 
$p(s)$ plots (Figs.~\ref{Real/p(s)}(a) and ~\ref{Real/p(s)}(c)). For 
IC2391 the $\bar{s}$ value and the $p(s)$ plot (Fig.~\ref{Real/p(s)}(b)) 
are most like those for clusters of type 3D1.

\subsection {The effect of subclusters on $p(s)$ and $\bar{s}$}

Chamaeleon and Taurus have correlation lengths intermediate between 
types 3D1 and 3D2, but their $p(s)$ plots are clearly not generic. 
This is because they contain sub-clusters, as illustrated in 
Figs. ~\ref{RealField}(d) and ~\ref{RealField}(e). 
Consequently $p(s)$ has multiple maxima. In some cases these maxima 
can be identified with (i) separations between stars in the same 
sub-cluster (the maximum at the smallest separations) and (ii) separations 
between stars in two distinct sub-clusters (maxima at larger separations, 
corresponding to the separation between the two sub-clusters). If there 
are ${\cal N}_{\rm sub}$ sub-clusters, there can be up to $1+{\cal N}_{\rm sub} 
({\cal N}_{\rm sub}-1)/2$ maxima, but fewer if there is degeneracy in the 
distances between sub-clusters. After smoothing, the $p(s)$ plot for 
Chamaeleon (Fig.~\ref{Real/p(s)}(e)) shows two distinct maxima, suggesting 
at least two sub-clusters, and Fig.~\ref{RealField}(e) does indeed show 
two sub-clusters. They are separated by $\sim 1$, hence giving rise to 
the maximum in $p(s)$ at $s \sim 1$. However, after smoothing, the $p(s)$ 
plot for Taurus (Fig.~\ref{Real/p(s)}(d)) shows only three well defined 
maxima, suggesting at most three sub-clusters, whereas  Fig.~\ref{RealField}(d) 
shows at least eight well defined sub-clusters. Evidently the $p(s)$ 
plot is not a robust diagnostic of sub-clustering.
 
If we now consider artificial fractal star clusters with the same  fractal
dimension D, we find that there is so much variance in their individual $\bar{p}(s)$
plots that we cannot sensibly define a mean $\bar{p}(s)$ plot. However, we can 
still compute the mean Normalised Correlation Length $\bar{s}$ and its variance.
The results are given in Table 1. We see that $\bar{s}$ increases monotonically
with increasing D and can therefore be used to estimate D for star clusters which 
are presumed {\it a priori} to be fractal.

Moreover, the value of $\bar{s}$ for star 
clusters of type F3.0 is essentially the same as for clusters of type 3D0, as
it should be. The small difference is attributable to the fact that in 
constructing clusters of type F3.0 the positioning of the individual stars is 
not completely random, whereas for  type 3D0 it is.

However, the range of $\bar{s}$
for D in (1.5,3.0) is almost identical to that for $\alpha$ in (0,2). Therefore
$\bar{s}$ is degenerate and cannot on its own be used to distinguish multi-scale 
(fractal) sub-clustering from a large-scale radial density gradient.

\section {Minimal Spanning Trees}

\begin{figure*}
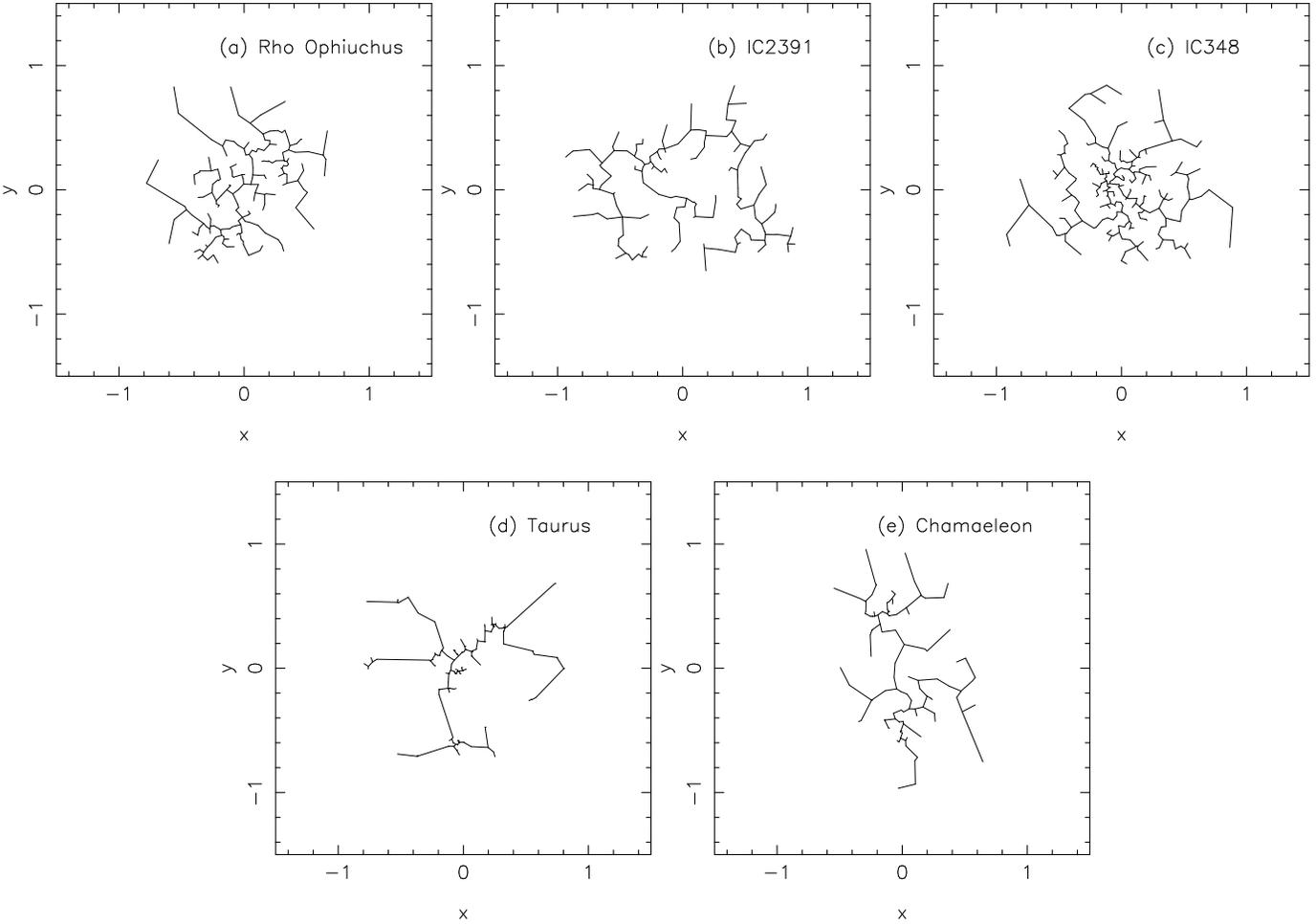

\centerline{\psfig{figure=fig5a.ps,width=6cm}
            \psfig{figure=fig5b.ps,width=6cm}
  	    \psfig{figure=fig5e.ps,width=6cm}}
\vspace{.5cm}
\centerline{\psfig{figure=fig5d.ps,width=6cm}
            \psfig{figure=fig5c.ps,width=6cm}} 
\caption{Minimal Spanning Trees for (a) $\rho$ Ophiuchus, (b) IC2391, 
(c) IC348, (d) Taurus, and (e) Chamaeleon.}
\label{MSTReal}
\end{figure*}

The Minimal Spanning Tree (MST) is the unique\footnote{ strictly speaking, 
if the array of points contains two or more pairs with exactly the same 
separation, the network may not be unique, as the points may be connected 
in a different order. However, even if this is the case, the total length 
of edges and the distribution of edge-lengths is preserved for all solutions.} 
network of straight lines joining  a set of points, such that the total 
length of all the lines -- hereafter `edges' -- in the network is minimised 
and there are no closed loops. The construction of such a tree is described 
by Gower and Ross (1969). Starting at any point, an edge is created joining 
that point to its nearest neighbour. The tree is then extended by always 
constructing the shortest link between one of its nodes and an unconnected  
point, until all the points have been connected. Figure~\ref{MSTReal} shows 
the MSTs for the real star clusters $\rho$ Ophiuchus, Taurus, Chamaeleon, 
IC348 and IC2391.

 
The use of Minimal Spanning Trees (MSTs) as a probe of cosmological structure 
was explored by Barrow, Bhavsar and Sonoda (1985), and a further 
refinement, the self avoiding random walk, was described by Baugh (1993). 
Although the approach seemed promising as a means of picking out clumps and 
filaments, the only statistical analysis of the MST for cosmological purposes
of which we are aware is due to Graham, Clowes and Campusano (1995), who
adopted methods developed by Hoffman and Jain (1983) and Dussert et al(1987)
and applied them to the distribution of quasars. We describe their analysis 
in Appendix~\ref{Graham}, and show that it does not work well for star clusters.

\subsection {The Normalised Mean Edge Length}
 
Once the MST of a cluster has been constructed it is straightforward to 
compute the Mean Edge Length, $\bar{m}$. Unlike the Normalised Correlation 
Length $\bar{s}$, $\bar{m}$ is not independent of the number of stars in 
the cluster, ${\cal N}_{\rm total}$. As ${\cal N}_{\rm total}$ increases, 
more short edges are created on the MST and $\bar{m}$ decreases. The 
expected total length of the MST of a random array of 
${\cal N}_{\rm total}$ points, uniformly distributed over a 
two-dimensional area A, is asymptotically proportional to 
$({\cal N}_{\rm total}A)^{1/2}$ (Hammersley et al. 1959). As 
there are ${\cal N}_{\rm total}-1$ edges, the mean edge length 
is asymptotically proportional to $({\cal N}_{\rm total}A)^{1/2} 
/({\cal N}_{\rm total}-1)$, and so this factor should be used to 
normalise the mean edge length of clusters having different areas 
$A$ and/or different numbers of stars ${\cal N}_{\rm total}$.

The resulting Normalised Mean Edge Length $\bar{m}$ has been evaluated 
for 100 realisations of each type of artificial star clusters, and for 
the real star clusters, and the results are recorded in Table 1 (column 
3). Also recorded in Table 1 (column 4) is the mean of the standard 
deviations of the MST edge lengths, $\bar{\sigma}_m$, This quantity 
is used in Appendix~\ref{Graham}.

\subsection{${\cal Q}$}

Table~\ref{Measures} shows that for artificial clusters of type 
2D$\alpha$, 3D$\alpha$ and $FD$, both $\bar{m}$ and $\bar{s}$ 
decrease monotonically as $\alpha$ increases (i.e. the degree of 
central concentration becomes more severe) or as $D$ decreases 
(i.e. the degree of sub-clustering become more severe). However, 
$\bar{s}$ decreases more quickly than $\bar{m}$ as $\alpha$ is 
increased, while $\bar{m}$ decreases more quickly than $\bar{s}$ 
as $D$ is decreased. Thus, the ratio 
\begin{equation}
{\cal Q} = \frac{\bar{m}}{\bar{s}}
\end{equation}
yields a measure which not only quantifies, but also distinguishes 
between, a smooth overall radial density gradient and multi-scale 
fractal sub-clustering. 

Mean values of ${\cal Q}$ for the various types of artificial star 
cluster are recorded in Table 1 (column 5). For artificial clusters 
with a smooth large-scale radial density gradient (type 3D$\alpha$), 
$\bar{\cal Q}$ increases from $\bar{\cal Q} \simeq 0.80$ to 
$\bar{\cal Q} \simeq 1.50$ as the degree of central concentration 
increases from $\alpha = 0$ (statistically uniform number-density) 
to $\alpha = 2.9$ ($n \propto r^{-2.9}$). For 
artificial clusters with fractal sub-structure (type F$D$), 
$\bar{\cal Q}$ decreases from $\bar{\cal Q} \simeq 0.80$ to 
$\bar{\cal Q} \simeq 0.45$ as the degree of sub-clustering increases 
from $D = 3.0$ (uniform number-density, no sub-clustering) to 
$D = 1.5$ (strong sub-clustering).

\begin{figure*}
\centerline{\psfig{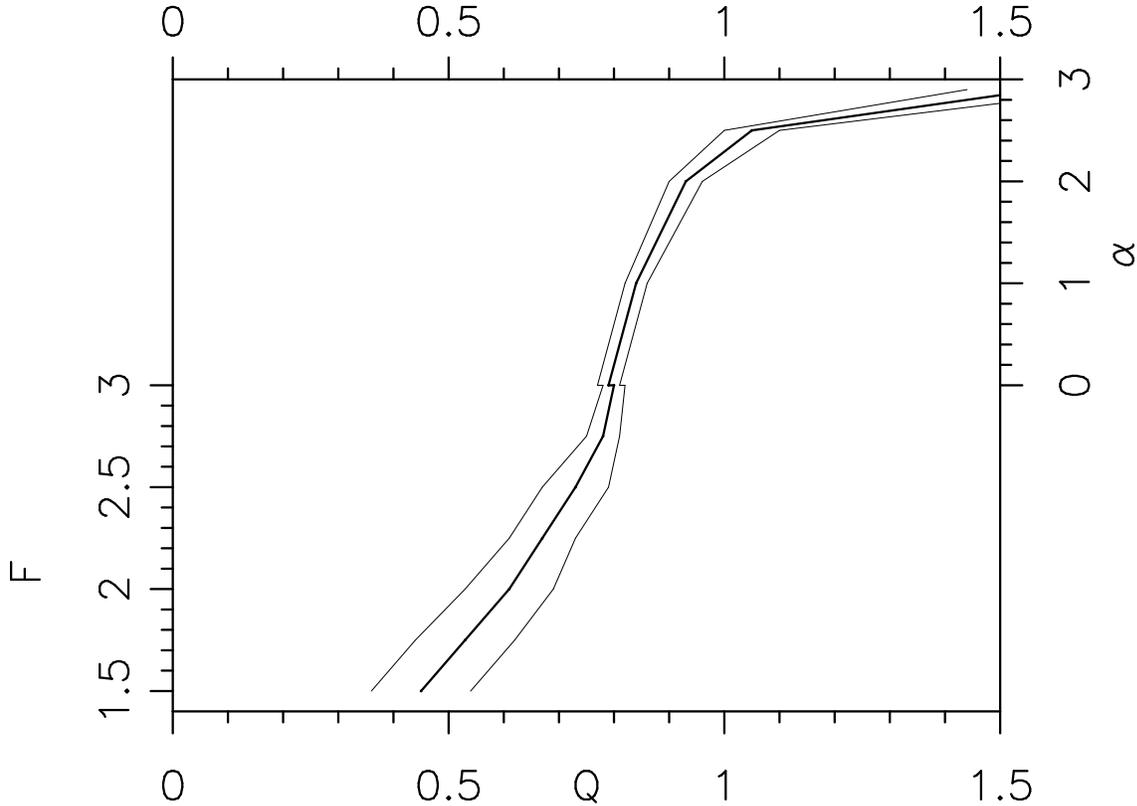}}
\caption{${\cal Q}$ plot for artificial star clusters. For ${\cal Q} \leq 0.80$, 
the fractal dimension $D$ should be read from the lefthand axis, and for 
${\cal Q} \geq 0.80$, the radial density exponent $\alpha$ should be read 
from the righthand axis. The small kink at ${\cal Q} \simeq 0.8$ is explained 
in the text.}
\label{Qplot}
\end{figure*}

We can therefore construct a plot (Figure ~\ref{Qplot}) of $D$ against 
${\cal Q}$ for ${\cal Q} \leq 0.80$, and $\alpha$ against ${\cal Q}$ 
for ${\cal Q} \geq 0.80$. For any real cluster we can compute its 
${\cal Q}$ value, and then use Figure ~\ref{Qplot} to read off its 
notional fractal dimension $D'$ (if ${\cal Q} < 0.80$, implying sub-clustering), 
or its radial density exponent $\alpha$ (if ${\cal Q} > 0.80$, implying 
a large-scale radial density gradient).

The small kink at ${\cal Q} \simeq 
0.8$ is due to the fact in constructing a cluster of type F3.0 the stars 
are positioned regularly (in the sense that at each generation, each 
subcube of space is occupied) and therefore the number-density is 
artificially uniform; in contrast, when we construct a cluster of type 
3D0 the stars are positioned randomly, so that the density is only uniform 
in a statistical sense and there are Poisson fluctuations in the local 
density.

Fractal dimensions obtained from Fig.~\ref{Qplot} in this way are only 
notional, because ${\cal Q}$ (or any other single measure) can reflect 
sub-clustering, but cannot capture whether the sub-clustering is 
hierarchically self-similar.

 
\begin{figure*}
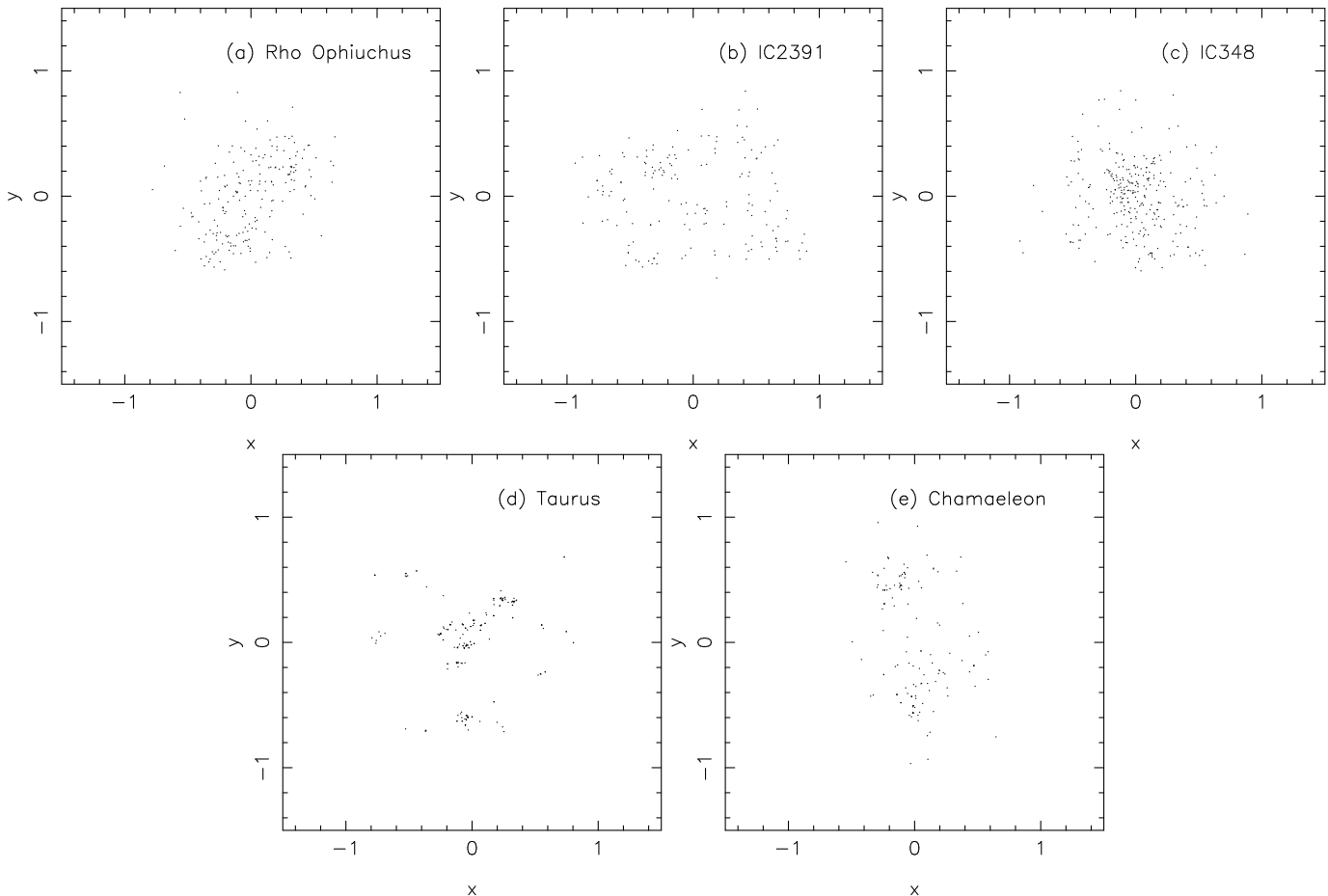

\centerline{\psfig{figure=figrho.ps,width=6cm}
            \psfig{figure=figic2391.ps,width=6cm}
            \psfig{figure=figic348.ps,width=6cm}}
\centerline{\psfig{figure=figtaur.ps,width=6cm}
            \psfig{figure=figcham.ps,width=6cm}}
\caption{Raw data for all real star clusters analysed in the paper. The 
clusters have been centred on the mean position of all stars and scaled 
so that the distance from the centre to the most distant star is unity.}
\label{RealField}
\end{figure*}

Using Figure ~\ref{Qplot}, we infer that Taurus, IC2391 and Chamaeleon 
have substructure with notional fractal dimensions 
$D = 1.5,\,2.2\,{\rm and}\,2.25\,$. In contrast, $\rho$ Ophiuchus and 
IC348 appear to be centrally concentrated, with radial density exponents 
$\alpha = 1.2\,{\rm and}\,2.2\,$. These inferences agree well with an 
intuitive reading of the raw data shown in Fig.~\ref{RealField}.

\subsection{The effect of binary companions on the MST Edge Length}

The MST will normally link a star to its binary companion, as this will 
usually be the shortest way of adding one or other of the stars to the 
tree. Binaries create very short edges and therefore a large population 
of binary stars will cause a noticeable reduction in the mean edge 
length, $\bar{m}$. Of the five real clusters considered in this paper, 
Taurus has been subjected to particularly close scrutiny and has a 
larger identified population of binaries than any of the others. As the 
binaries are not part of the clustering regime, it is important to 
establish whether they are distorting the result.

Using the MST, all pairs of stars lying closer together than $10^{-4}$ 
of the cluster radius were pruned, leaving single stars. For Chamaeleon, 
$\rho$ Ophiuchus, IC348 and IC2391, only 3, 0, 1 and 2 such pairs were 
found; Taurus, however, was pruned from 215 down to only 137 primary 
stars. For the pruned version of Taurus, $\bar{s}$ increased from 0.55 
to 0.57, while $\bar{m}$ increased from 0.26 to 0.33 and ${\cal Q}$ 
increased from 0.47 to 0.58. Removal of the binaries resulted in the 
notional fractal dimension for Taurus being increased from 1.5 to 1.9. 
This demonstrates that in a cluster with a large binary population, it 
is important to prune the close companions before evaluating $\bar{s}$, 
$\bar{m}$ and ${\cal Q}$.


\begin{table}
\caption{Sources of positions for cluster members and approximate ages and
crossing times for clusters. (Crossing times were calculated using a typical
velocity dispersion of 2 km/sec.) }
\begin {tabular}  {lcccl}\hline
Name &  Members& Age&$ T_{\rm cross}$ & Sources \\
 & & Myr& Myr &\\
 & & & & \\
IC2391 & 166 &53&2.5 & Barrado et al. (2001) \\
Cham. & 136 & 0.1-40 &2.7&Lawson et al. (1996) \\
 & & & & Ghez et al. (1997) \\
Taurus & 215 & 1.0 & 10.0 & Briceno et al. (1993) \\
 & & & & Ghez et al. (1993) \\
 & & & & Gomez et al. (1992) \\
 & & & & Hartmann et al. (1991)\\
 & & & & Herbig et al. (1988) \\
 & & & & Leinert et al. (1993) \\
 & & & & Simon et al. (1995) \\
 & & & & Waer et al. (1988) \\
 & & & & Luhman et al. (2003) \\
$\rho$ Oph. & 199 &0.3 - 2.0 & 1.35& Bontemps et al. (2001) \\ 
IC348 & 288 & 2.0 &2.0& Luhman et al. (2003) \\ \hline
\end {tabular}
\label{Sources}
\end {table}



\begin{figure*}
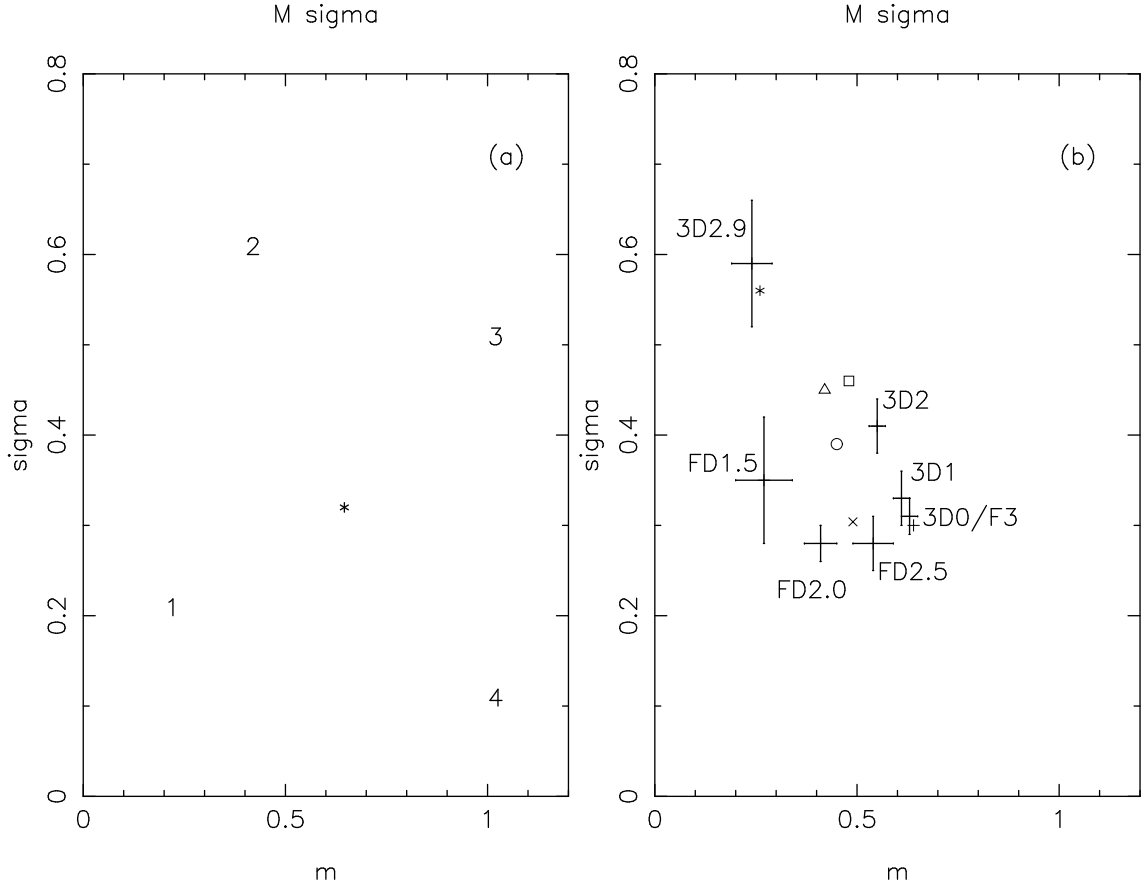

\centerline{\psfig{figure=figmsig1.ps,width=7.5cm}
            \psfig{figure=figmsig2.ps,width=7.5cm}}
            
\caption{MST $(m,\sigma_m)$-plots.  (a) $(m,\sigma_m)$-plane, showing the 
regions of the plane in which well characterised distributions of points 
converge (from Graham et al). *: random distributions, 1: clustered 
structures, 2: concentration gradients, 3: quasi periodic tilings, 4: 
highly organised distributions. (b) $(m,\sigma_m)$-plane, with artificial 
star clusters of types 3D0 to 3D2.9 and F1.5 to F3.0 plotted. The five real 
clusters are indicated by symbols $\ast$ : Taurus, $\circ$ : $\rho$ Ophiuchus, 
$\times$ : IC2391,  $\Box$ : IC348, $\triangle$ : Chamaeleon.}
\label{mstbar}
\end{figure*}
 

\section{Discussion}\label{Discussion}

The ratio of the Normalized Mean Edge Length to the Normalized 
Correlation Length, ${\cal Q}$, is effective in distinguishing between 
a smooth large-scale radial density gradient and multi-scale fractal 
sub-clustering, because it is sensitive not only to the frequency of 
small separations between stars, but also to their spatial distribution.

The MST Edge length  $\bar{m}$ is a simple average of the distances of 
stars to their (usually) closest neighbours. If one star is moved very 
close to another, the change in $\bar{m}$ will be diluted by the total 
number of stars, ${\cal N}_{\rm total}$. However, when calculating the 
mean distance of companions for all stars, only one star out of 
${\cal N}_{\rm total}$ in the cluster has had one of its 
${\cal N}_{\rm total} -1$ companions moved very close. The change in 
$\bar{s}$, is therefore diluted by ${\cal N}_{\rm total} ^ {2}$. Thus, 
when small separations are scattered all over the cluster, increasing 
the number of small separations causes both  $\bar{m}$ and  $\bar{s}$ to 
decrease,but  $\bar{s}$ decreases more slowly than $\bar{m}$. This is 
the case for decreasing fractal dimension.

For radially concentrated clusters, by contrast, increasing the 
clustering creates more small separations between stars, but these are 
all in the central region of the cluster. Moving another star to this 
area affects $\bar{m}$ in the normal way, the star having a newly short 
edge length between it and its nearest neighbour, and the change in the 
mean distance being diluted by ${\cal N}_{\rm total}$. However, the 
large number of other stars in the centre also gain another close 
neighbour. The decrease in $\bar{s}$ is therefore compounded and 
exceeds that in $\bar{m}$.

Consequently, the quotient ${\cal Q} = \bar{m} / \bar{s}$ successfully 
distinguishes between clusters which have a smooth large-scale radial 
density gradient and clusters which have multi-scale fractal 
sub-clustering, in a way which agrees with an intuititive analysis but 
which cannot be accomplished using existing methods such as Larson 
Plots or Box Dimension Plots. An additional advantage over these 
methods is that the calculation of ${\cal Q}$ is quantitative and 
objective, as no intervention is required in the normalisation process, 
in the construction of the MST, or in choosing a range over which to 
calculate a slope.

We should emphasize that classical methods for evaluating the density 
profile of a cluster, or its fractal dimension, are not viable for 
clusters with $\sim$ 200 members, primarily because of low-number statistics. 
For example, if one attempts to define the projected radial density 
profile for a real cluster of stars by counting stars in different 
annuli, the result is very noisy.

Alternatively, if one attempts to 
determine the mean projected radial density profile for a 200-member 
artificial cluster having a given radial density profile in three 
dimensions, using many different realizations and with a view to 
comparing this with a real cluster, one finds that the standard deviation 
is very large, and so the diagnostic power of this profile is poor. 

In the same spirit one might attempt to construct the Box Dimension Plot
(BDP) of a real cluster and compare it with the mean BDP of artificial star
clusters having a given fractal dimension. To construct a BDP one divides
the projected image of the star cluster into a grid of square cells of side 
$l$ and counts the number of cells, ${\cal N}_{\rm occ}(l)$ which are occupied 
by stars.
Then, by repeating this for different values of $l$, one obtains a plot of 
$log({\cal N}_{\rm occ}(l))$ against $-log(l)$. For a true fractal this plot is a straight
line with slope equal to the fractal dimension. However, for a star cluster 
with only  $\sim 200 $ members, the plot is not linear. By treating many realisations of artificial
clusters all having the same fractal dimension {\em and the same number of 
stars}, one can define a mean BDP. 
However, the mean BDP is not very strongly dependent
on the fractal dimension and it has a large standard deviation. Therefore the
 Box Dimension Plot of a real cluster does not give a useful constraint on
  its fractal dimension. 

It is for this reason that we have sought integral measures of cluster 
structure. The same philosophy informs the use of equivalent width when 
evaluating noisy spectral lines (for example).

We also note that a cluster cannot have a large-scale radial density 
gradient, and at the same time be fractally sub-clustered. A cluster 
could have a large-scale radial density gradient and {\em non-fractal} 
sub-clustering -- but then it would require more parameters to 
characterize the structure, and its diagnosis would become 
correspondingly more difficult (if not impossible for clusters with 
$\sim$ 200 stars).

In Table 2 we list estimates for the ages and the crossing times of the clusters
we have analyzed. On the basis of simple arguments, we might expect the 
${\cal Q}$ value of a cluster to increase with time, as the substructure 
dissolves and the overall cluster relaxes to a radially concentrated 
density profile. However, this is not evident in the small sample treated here.
Taurus has an age much less than its crossing time, which is consistent with its 
small ${\cal Q}$ value and low fractal dimension. On the other hand, IC2391 
and Chamaeleon have ages much greater than their crossing times and yet they are
still fractal with relatively low ${\cal Q}$ values. In contrast,  $\rho$ 
Ophiuchus and IC348, which have ages comparable with their crossing times,
are both centrally condensed, with no discernible substructure. We should, 
however, caution against drawing firm conclusions from such a small sample.
We also note that young clusters observed at short wavelengths (i.e. in the 
optical), may appear to have substructure due to patchy observation. therefore
long wavelength surveys are preferable for embedded young star clusters.

\section{Conclusions}\label{Conclusions}

We have explored two statistical measures for analysing objectively 
the observed (i.e. projected) structures of star clusters. These 
measures are based on the Mean Surface Density of Companions (MSDC), 
and the Minimal Spanning Tree (MST). The measures are $\bar{s}$, 
the normalised mean separation between stars, and $\bar{m}$, the 
normalised mean edge-length of the MST, both of which are independent 
of the number of stars in the cluster. For artificial star clusters, 
created with a smooth large-scale radial density profile 
($n \propto r^{-\alpha}$), and for artificial star clusters created with 
sub-structure having fractal dimension $D$, $\bar{s}$ and $\bar{m}$ 
both decrease with increasing $\alpha$ and/or decreasing $D$ -- but 
at different rates. Hence a cluster with a radial gradient can be 
distinguished from one with sub-structure by evaluating 
${\cal Q} = \bar{m} / \bar{s}$. For a cluster of  uniform 
volume-density (i.e. $\alpha = 0$ and $D = 3.0$), ${\cal Q} \simeq 0.80$. 
If the cluster is made more centrally condensed by increasing $\alpha$, 
${\cal Q}$ increases monotonically, reaching ${\cal Q} \simeq 1.50$ at 
$\alpha = 2.9$. Conversely, if the cluster is given sub-structure by 
reducing $D$, ${\cal Q}$ decreases monotonically, reaching 
${\cal Q} \simeq 0.45$ at $D = 1.5$.

On the basis of their ${\cal Q}$ values, $\rho$ Ophiuchus and IC348 
have radial gradients with $\alpha \simeq 1.2 \pm 0.3$, and 
$2.2 \pm 0.2$, respectively. Chamaeleon and IC2391 have sub-structure 
with notional fractal dimension $D' \simeq 2.2 \pm 0.2$. Taurus has 
even more sub-structure, with $D' \simeq 1.55 \pm 0.25$, and if the 
binaries in Taurus are treated as single systems, $D'$ increases to 
$1.9 \pm 0.2$. $D'$ is only a notional fractal dimension, because the 
integral measures we have defined do not give any indication of whether 
the sub-structure is hierarchically self-similar. (Indeed, for clusters 
having only $\sim$ 200 stars the range of separations is too small to 
possess hierarchical self-similarity.)



\appendix
\section{Raw Data}\label{RawData}

Table~\ref{Sources} gives the sources of the positions of stars --- or, in 
the case of $\rho$ Ophiuchus, protostars --- used in the analysis of Sections 
3 and 4. These positions are plotted on Fig.~\ref{RealField}.

\section{Detecting randomness and clustering using the Minimal 
Spanning Tree.}\label{Graham}

Graham et al (1995) have applied the MST to quasar clustering on very 
large scales, using a method which was developed by Dussert et al (1987) 
for characterising biological structures. In Dussert's method, the mean 
$\bar{m}$ and standard deviation $\sigma_m$ of the edge lengths of the 
MST are first computed and normalised by dividing by the factor 
$({\cal N}_{\rm total}A)^{1/2}/({\cal N}_{\rm total}-1)$, then plotted 
on the $(m,\sigma_m)$-plane.  
Fig.~\ref{mstbar}(a), reproduced from Dussert et al. (1987), shows 
the theoretical locations on the $(m,\sigma_m)$-plane for different 
types of clustering in two-dimensions. The region of the 
$(m,\sigma_m)$-plane around the central star represents the locus of 
a random distribution. The region of the $(m,\sigma_m)$-plane around 
`1' represents the locus of distributions dominated by sub-clustering. 
The region of the $(m,\sigma_m)$-plane around `2' represents the 
locus of distributions dominated by radial concentration gradients. 
The region of the $(m,\sigma_m)$-plane around `3' represents the 
locus of distributions dominated by quasi-periodic tilings. And the 
region of the $(m,\sigma_m)$-plane around `4' represents the locus 
of highly organized distributions (i.e. lattices).

Fig.~\ref{mstbar}(b) shows the loci on the $(m,\sigma_m)$-plane for 
the various artificial star cluster types and the five real clusters, 
and reveals some drawbacks to this plot. The locus for artificial 
clusters with a radial density gradient do indeed tend towards region 
2 with increasing $\alpha$ (i.e. greater degree of central concentration), 
although only for $\alpha \ga 2$ are they clearly distinguishable 
from a purely random distribution. Similarly, the locus for artificial 
clusters with fractal sub-clustering tend towards region 1 with decreasing 
$D$ (i.e. greater degree of sub-clustering) for $D \ga 2.0$. However, for 
$D \la 2.0$, this trend is abandoned, and the locus moves towards region 
2; in other words, a cluster with a low fractal dimension and hence a 
high degree of sub-clustering masquerades -- on the $(m,\sigma_m)$-plane 
-- as a cluster with a strong radial density gradient, albeit it not 
precisely of the form $n \propto r^{-\alpha}$. Moreover, Taurus, which to 
the human eye appears to have the most well defined sub-clustering of 
all five real clusters, masquerades on the $(m,\sigma_m)$-plane as a 
cluster with a strong radial density gradient, $\alpha \simeq 2.7$.

We conclude that the $(m,\sigma_m)$-plane is not able to distinguish 
between a smooth large-scale radial density gradient and multi-scale 
fractal sub-clustering

\end{document}